\PassOptionsToPackage{hidelinks, colorlinks=true, allcolors=linkcolor}{hyperref}
\documentclass[submission,copyright,creativecommons]{eptcs}
\providecommand{\event}{ICLP 2025} 

\usepackage{iftex}

\ifpdf
  \usepackage[T1]{fontenc}        
\else
  \usepackage{breakurl}           
\fi

\usepackage{url}
\usepackage{mathptmx}
\usepackage{booktabs}
\usepackage{dsfont}
\usepackage{subcaption}
\usepackage{wrapfig} 
\usepackage{amssymb}
\usepackage{amsmath}
\usepackage{graphicx}
\usepackage{multirow}
\usepackage{multicol}
\usepackage{xspace}
\usepackage{xcolor}
\definecolor{linkcolor}{HTML}{A93C93}
\usepackage{url}

\usepackage[
textwidth=0.125\textwidth,textsize=footnotesize]{todonotes}
\iffalse 
    \newcommand{\comm}[1]{\todo[bordercolor=yellow,linecolor=yellow]{\textsf{\scriptsize #1}}}
    \newcommand{\commin}[1]{\todo[inline,bordercolor=yellow,linecolor=yellow]{\textsf{\scriptsize #1}}}

    \newcommand{\MYTODO}  [1]{\textcolor{noteColor1}{\textbf{\small[[TODO: #1]]}}\\}
    \newcommand{\mytodo}  [1]{\textcolor{noteColor1}{\textbf{\small[[TODO: #1]]}}} 
    \newcommand{\mytodoF}  [1]{\textcolor{noteColor1}{\textbf{[[TODO: #1]]}}} 
    
\else
    \newcommand{\comm}[1]{}
    \newcommand{\commin}[1]{}
    
    \newcommand{\MYTODO}  [1]{}
    \newcommand{\mytodo}  [1]{}
    \newcommand{\mytodoF}  [1]{}
    
\fi

\usepackage{listings}

\usepackage[most]{tcolorbox}
\usepackage{enumitem}

\usepackage{tikz}
\newcommand{\linkIcon}{%
    \tikz[x=1.2ex, y=1.2ex, baseline=-0.05ex]{%
        \begin{scope}[x=1ex, y=1ex]
            \clip (-0.1,-0.1) 
                --++ (-0, 1.2) 
                --++ (0.6, 0) 
                --++ (0, -0.6) 
                --++ (0.6, 0) 
                --++ (0, -1);
            \path[draw, 
                line width = 0.5, 
                rounded corners=0.5] 
                (0,0) rectangle (1,1);
        \end{scope}
        \path[draw, line width = 0.5] (0.5, 0.5) 
            -- (1, 1);
        \path[draw, line width = 0.5] (0.6, 1) 
            -- (1, 1) -- (1, 0.6);
        }%
    }

\definecolor{noteColor1} {rgb}{  1,  0,  0} %
\definecolor{noteColor2} {rgb}{  1,0.5,  0} %
\definecolor{noteColor3} {rgb}{0.8,0.8,  0}
\definecolor{noteColor4} {rgb}{  1,  0,0.6}
\definecolor{noteColor5} {rgb}{  1,  0,  1} %
\definecolor{noteColor6} {rgb}{  0,0.8,  0} %
\definecolor{noteColor7} {rgb}{  0,0.8,0.6} %
\definecolor{noteColor8} {rgb}{  0,0.8,0.8} %
\definecolor{noteColor9} {rgb}{0.5,0.8,  0} %
\definecolor{noteColor10}{rgb}{  0,  0,  1} %
\definecolor{noteColor11}{rgb}{0.5,  0,  1} %
\definecolor{noteColor12}{rgb}{  0,0.5,  1} %
\definecolor{noteColor13}{rgb}{0.5,0.5,0.5}
\definecolor{maroon}{rgb}{0.5,0,0}
\definecolor{darkgreen}{rgb}{0,0.5,0}
\definecolor{mildgreen}{rgb}{0,0.8,0}
\definecolor{lightblue}{rgb}{0.3,0.3,1}
\definecolor{lightgrey}{rgb}{0.97,0.97,0.97}
\definecolor{grey}{rgb}{0.65,0.65,0.65}
\definecolor{codebg}{rgb}{0.97,0.97,0.97}

\lstnewenvironment{lstlistingwolinenum}{
    \lstset{
        basicstyle=\ttfamily\footnotesize,
        frame=lrtb,
        breaklines=true
    }
}{}
\lstnewenvironment{lstlistingwlinenum}{
    \lstset{
        basicstyle=\ttfamily\footnotesize,
        frame=lrtb,
        breaklines=true,
        numbers=left,
        numbersep=0.5em,
        xleftmargin=1.5em,
        framexleftmargin=1.5em,
        stepnumber=1
    }
}{}

\usepackage{ tipa }
\usepackage{adjustbox}
\usepackage{relsize}
\definecolor{PrologPredicate}{RGB}{0,0,200}
\definecolor{PrologVar}      {RGB}{105,0,175}
\definecolor{PrologComment}  {RGB}{169,082,044}
\definecolor{PrologOther}    {rgb}{0.2,0.2,0.2}
\definecolor{PrologString}   {RGB}{070,120,200}
\usepackage{listings} 
\usepackage{textcomp}
\newcommand{\code}{\lstinline[style=MyInline]}
\lstdefinestyle{MyInline}
{
  basicstyle = \relsize{-0.5}\ttfamily\color{PrologPredicate},
  escapechar = @,
  escapeinside = {-<}{>-},
  breaklines = true,
  breakatwhitespace=true,
  upquote = true,
  literate =
  {?-}{{?-\,}}3
  {:-}{{:-\,}}3
  {.=.}{{\,\#=\,}}3
  {.<.}{{\,\#<\,}}3
  {.>.}{{\,\#>\,}}3
  {.=<.}{{\,\#=<\,}}4
  {.>=.}{{\,\#>=\,}}4
}
\lstdefinestyle{MySCASP}
{
    showlines=true,
     numbersep=1em,    
     xleftmargin=0.35cm,  
    numberstyle=\tiny,
    numbers=left,
    stepnumber=1,
  breaklines = false,
  mathescape = true,
  escapechar = @,
  escapeinside = {-<}{>-},
  keywords = {},
  upquote = true,
  basicstyle = \linespread{0.75}\ttfamily\color{PrologPredicate}\footnotesize,  
  basewidth = 0.43em,
  moredelim = {*[s][\color{PrologVar}]{(}{)}},
  moredelim = {*[s][\color{PrologString}]{'}{'}},
  moredelim = {*[s][\color{PrologOther}]{:-}{.}},
  commentstyle = \mdseries\color{PrologComment},
  escapebegin=\color{PrologVar},
  morecomment=[l]\%,
  literate     =
  {|}{{|}}2
  {&(}{{(}}1
  {&)}{{)}}1
  {.=.}{{\,\#=\,}}2
  {.<.}{{\,\#<\,}}2
  {.>.}{{\,\#>\,}}2
  {.<>.}{{\,\#\textdoublebarslash\,}}2
  {.=<.}{{\,\#=<\,}}3
  {.>=.}{{\,\#>=\,}}3
}

\usepackage{ntheorem}

{
  \theoremindent=.5cm
  \theoremheaderfont{\kern-.5cm\normalfont\bfseries}
  \theorembodyfont{\normalfont}
  \theoremseparator{:}
  \theoremprework{\setlength\parindent{0cm}}
  \newtheorem{example}{Example}
}

{
  \theoremindent=.5cm
  \theoremheaderfont{\kern-.5cm\normalfont\bfseries}
  \theorembodyfont{\normalfont}
  \theoremseparator{:}
  \theoremprework{\setlength\parindent{0cm}}
  \newtheorem{apxExample}{Example}
}

\usepackage{caption}
\newcommand{\version}{0.25.06.25\xspace}

\newcommand{\secVspaceB}{\vspace{-1em}}         
\newcommand{\secVspaceA}{\vspace{-0.5em}}       

\newcommand{\ssecVspaceB}{\vspace{-0.8em}}        
\newcommand{\ssecVspaceA}{\vspace{-0.3em}}      

\newcommand{\sssecVspaceB}{\vspace{-0.8em}}       
\newcommand{\sssecVspaceA}{\vspace{-0.3em}}     

\newcommand{\figVspaceb}{\vspace{-0.5em}}       
\newcommand{\figVspaceA}{\vspace{-0.5em}}       
\newcommand{\tabVspaceA}{\vspace{-1.8em}}       

\newcommand{\artifactDoi}{\raisebox{-0.2em}{\href{https://doi.org/10.5281/zenodo.15737913}{%
\includegraphics[height=1em]{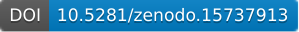}%
}}}

\newcommand{\repoExamples}{\url{https://github.com/ovasicek/iclp25-scasp-zeno/tree/master/examples}}
\newcommand{\myurl}[1]{\href{https://github.com/ovasicek/iclp25-scasp-zeno/blob/master/examples/#1}{\nolinkurl{#1}}}
\newcommand{\myurlwline}[2]{\href{https://github.com/ovasicek/iclp25-scasp-zeno/blob/master/examples/#1\#L#2}{\nolinkurl{#1}}, line #2}

\newcommand{\myurlExOne     }[1]{\href{https://github.com/ovasicek/iclp25-scasp-zeno/blob/master/examples/ex1-light_on_off/#1}{\nolinkurl{ex1/#1}}}
\newcommand{\myurlExTwo     }[1]{\href{https://github.com/ovasicek/iclp25-scasp-zeno/blob/master/examples/ex2-simplified_bank_account/#1}{\nolinkurl{ex2/#1}}}
\newcommand{\myurlExThree   }[1]{\href{https://github.com/ovasicek/iclp25-scasp-zeno/blob/master/examples/ex3-fading_light/#1}{\nolinkurl{ex3/#1}}}
\newcommand{\myurlExFour    }[1]{\href{https://github.com/ovasicek/iclp25-scasp-zeno/blob/master/examples/ex4-pulsing_light/#1}{\nolinkurl{ex4/#1}}}
\newcommand{\myurlExFive    }[1]{\href{https://github.com/ovasicek/iclp25-scasp-zeno/blob/master/examples/ex5-simplified_water_tanks/#1}{\nolinkurl{ex5/#1}}}
\newcommand{\myurlExSix     }[1]{\href{https://github.com/ovasicek/iclp25-scasp-zeno/blob/master/examples/ex6-water_tanks/#1}{\nolinkurl{ex6/#1}}}

\newcommand{\myurlExApxOne }[1]{\href{https://github.com/ovasicek/iclp25-scasp-zeno/blob/master/examples/apx1-blinking_light/#1}{\nolinkurl{apx1/#1}}}

\newcommand{\myurlExOtherOne   }[1]{\href{https://github.com/ovasicek/iclp25-scasp-zeno/blob/master/examples/other/1-bank_account/#1}{\nolinkurl{other/1/#1}}}
\newcommand{\myurlExOtherTwo   }[1]{\href{https://github.com/ovasicek/iclp25-scasp-zeno/blob/master/examples/other/2-light_trigger_off/#1}{\nolinkurl{other/2/#1}}}
\newcommand{\myurlExOtherThree   }[1]{\href{https://github.com/ovasicek/iclp25-scasp-zeno/blob/master/examples/other/3-train_gate/#1}{\nolinkurl{other/3/#1}}}
\newcommand{\myurlExOtherFour }[1]{\href{https://github.com/ovasicek/iclp25-scasp-zeno/blob/master/examples/other/4-falling_object/#1}{\nolinkurl{other/4/#1}}}
\newcommand{\myurlExOtherFive  }[1]{\href{https://github.com/ovasicek/iclp25-scasp-zeno/blob/master/examples/other/5-filling_vessel/#1}{\nolinkurl{other/5/#1}}}
\newcommand{\myurlExOtherSix  }[1]{\href{https://github.com/ovasicek/iclp25-scasp-zeno/blob/master/examples/other/6-no_loss_water_tanks/#1}{\nolinkurl{other/6/#1}}}
\newcommand{\myurlExOtherSeven   }[1]{\href{https://github.com/ovasicek/iclp25-scasp-zeno/blob/master/examples/other/7-no_loss_bouncing_ball/#1}{\nolinkurl{other/7/#1}}}
\newcommand{\myurlExOtherEight }[1]{\href{https://github.com/ovasicek/iclp25-scasp-zeno/blob/master/examples/other/8-bouncing_ball/#1}{\nolinkurl{other/8/#1}}}

\newcommand{\linkIconExOne  }{\href{https://github.com/ovasicek/iclp25-scasp-zeno/blob/master/examples/ex1-light_on_off}{\linkIcon{}}}
\newcommand{\linkIconExTwo  }{\href{https://github.com/ovasicek/iclp25-scasp-zeno/blob/master/examples/ex2-simplified_bank_account}{\linkIcon{}}}
\newcommand{\linkIconExThree}{\href{https://github.com/ovasicek/iclp25-scasp-zeno/blob/master/examples/ex3-fading_light}{\linkIcon{}}}
\newcommand{\linkIconExFour }{\href{https://github.com/ovasicek/iclp25-scasp-zeno/blob/master/examples/ex4-pulsing_light}{\linkIcon{}}}
\newcommand{\linkIconExFive }{\href{https://github.com/ovasicek/iclp25-scasp-zeno/blob/master/examples/ex5-simplified_water_tanks}{\linkIcon{}}}
\newcommand{\linkIconExSix}{\href{https://github.com/ovasicek/iclp25-scasp-zeno/blob/master/examples/ex6-water_tanks}{\linkIcon{}}}

\newcommand{\linkIconExApxOne }{\href{https://github.com/ovasicek/iclp25-scasp-zeno/blob/master/examples/apx1-blinking_light}{\linkIcon{}}}

\newcommand{\linkIconExOtherOne   }{\href{https://github.com/ovasicek/iclp25-scasp-zeno/blob/master/examples/other/1-bank_account}{\linkIcon{}}}
\newcommand{\linkIconExOtherTwo   }{\href{https://github.com/ovasicek/iclp25-scasp-zeno/blob/master/examples/other/2-light_trigger_off}{\linkIcon{}}}
\newcommand{\linkIconExOtherThree   }{\href{https://github.com/ovasicek/iclp25-scasp-zeno/blob/master/examples/other/3-train_gate}{\linkIcon{}}}
\newcommand{\linkIconExOtherFour }{\href{https://github.com/ovasicek/iclp25-scasp-zeno/blob/master/examples/other/4-falling_object}{\linkIcon{}}}
\newcommand{\linkIconExOtherFive  }{\href{https://github.com/ovasicek/iclp25-scasp-zeno/blob/master/examples/other/5-filling_vessel}{\linkIcon{}}}
\newcommand{\linkIconExOtherSix  }{\href{https://github.com/ovasicek/iclp25-scasp-zeno/blob/master/examples/other/6-no_loss_water_tanks}{\linkIcon{}}}
\newcommand{\linkIconExOtherSeven   }{\href{https://github.com/ovasicek/iclp25-scasp-zeno/blob/master/examples/other/7-no_loss_bouncing_ball}{\linkIcon{}}}
\newcommand{\linkIconExOtherEight }{\href{https://github.com/ovasicek/iclp25-scasp-zeno/blob/master/examples/other/8-bouncing_ball}{\linkIcon{}}}

\title{
On Zeno-like Behaviors in the Event Calculus with Goal-directed Answer Set Programming%
\thanks{%
    \relsize{-0.5}%
     The work was supported by the project 23-06506S of the Czech Science
     Foundation and the FIT BUT internal project FIT-S-23-8151; grant VAE (TED2021-131295B-C33) funded by
     MCIN/AEI/10.13039/501100011033 and by the “European Union
     NextGenerationEU/PRTR”, grant COSASS (PID2021-123673OB-C32) funded by
     MCIN/AEI/10.13039/501100011033 and by “ERDF A way of making Europe”;
     by US NSF Grants IIS 1910131 and grants from industry through the UT Dallas Center 
     for Applied AI and Machine Learning.
}%
} 
\def\titlerunning{On Zeno-like Behaviors in EC with Goal-directed ASP}

\author{%
    Ond\v{r}ej Va\v{s}\'{i}\v{c}ek$^1$ \hfill Joaquin Arias$^2$ \hfill Jan Fiedor$^{1,6}$ \hfill Gopal Gupta$^3$ \hspace{.11cm} \\
    \hspace{.05cm} Brendan Hall$^4$ \hfill \hspace{.15cm} Bohuslav K\v{r}ena$^1$ \hfill Brian Larson$^5$ \hfill Tom\'{a}\v{s} Vojnar$^{1,7}$
    \institute{%
    $^1$Brno University of Technology, CZ \hspace{.3cm} $^2$CETINIA, Universidad Rey Juan Carlos, Spain \hspace{.3cm} $^3$UT Dallas, USA\\
    $^4$Ardent Innovation Labs, USA \hspace{.1cm} $^5$Multitude Corp., USA \hspace{.1cm} $^6$Honeywell International s.r.o., CZ \hspace{.1cm} $^7$Masaryk University, CZ
    }%
}%

\def\authorrunning{O. Va\v{s}\'{i}\v{c}ek et al.}

\begin{document}

\maketitle

\label{firstpage}

\begin{abstract}
It has been argued that Event Calculus (EC) is suitable for modeling high-level specifications of safety-critical cyber-physical systems.
The primary advantage lies in the rather small semantic gap between EC models and requirements expressed in a semi-formal natural language.
Moreover, its use of continuous time and variables avoids imprecision that stems from discretization.
In the past, we have shown that a goal-directed ASP system can be used for implementing these EC models.   
However, precise representation of time as an infinitesimally divisible continuous quantity leads to Zeno-like behaviors and to non-termination in such a system.
In this work, we model a number of well-known example problems from the literature to systematically study various natural EC modeling patterns that yield these Zeno-like behaviors, and
propose ways to deal with them. Moreover, we also propose a technique to automatically detect all such cases.
\end{abstract}



%

\secVspaceB{}\section{Introduction and Motivation}\secVspaceA{}\label{sec:intro}


Requirements engineering is a wide field of research due to its
various applications in critical sectors, such as medical or
aerospace, where it is necessary to verify the correct operation
of cyber-physical systems (CPS).
Their development starts with the specification of \emph{system requirements}, which is by many considered the most crucial part of the development process.
The importance is clearly illustrated by the results of the AVSI SAVI project~\cite{savi} which show that 70\% of errors are introduced 
during the system specification and design phases, yet most of them 
are detected later, during the testing of system's concrete implementation.
The main reason the errors are not detected earlier is that the system specification describes a~whole set of feasible solutions from which one concrete solution is 
eventually 
chosen to be implemented.
Such a single solution is much easier to verify compared to a potentially huge set of solutions to validate in the early phases
, mainly because we can make assumptions, e.g., that the CPS operates in discrete time.
When reasoning about all feasible solutions, we need to represent their real-world behavior as closely as possible which includes modeling the system behavior in continuous time and space.
However, this introduces 
issues such as Zeno behavior where, 
theoretically,
an infinite number of events 
can occur in a finite interval---an unrealistic behavior which makes modeling difficult.

As argued in~\cite{iclp24-pcapump-ec-scasp-shortened}, a promising way to reason about system requirements is to transform them into \emph{Event Calculus} (EC)~\cite{mueller_book-fixed} 
and use its semantics for the verification and validation.
EC is particularly suitable 
since its semantics follows how a human would think of the requirements, which makes the semantic gap between EC and the requirements near-zero.
This allows EC to faithfully reason about the behavior defined by the requirements without being tainted by 
design or implementation decisions, such as ``designing" states, transitions, or decomposition into sub-systems as is the case, e.g., with automata-based approaches.
The capabilities of this approach have already been shown on a train gate controller system~\cite{gupta-train-shortened} and in our prior work~\cite{iclp24-pcapump-ec-scasp-shortened}, where we managed to discover a~number of inconsistencies and a violation of a safety property in the specification of a safety-critical medical device.
Both of the above works use \emph{s(CASP)}---a goal-directed, grounding-free reasoner for Answer Set Programming (ASP) with constraints~\cite{scasp-iclp2018-shortened}.
The grounding-free nature and support for constraint solving allow s(CASP) to reason about \emph{continuous time} and \emph{continuous quantities}, which are both crucial for accurately and realistically representing CPS behavior.
The goal-directed nature further allows s(CASP) to provide \emph{explainable results}, which is crucial when using results of automated reasoning as evidence for certification, e.g., when building assurance cases~\cite{clarissa-assurance-shortened}.
However, similar to Prolog, s(CASP) can suffer from \emph{non-termination}, especially when reasoning in continuous time
, which both of the above works struggled~with.

In this work, we identify and systematically classify a number of general \emph{Zeno-like behaviors} which cause non-termination in goal-directed reasoning under continuous time.
We have implemented all relevant examples from the literature,
including fairly complex ones like the water tanks~\cite{97-alur-henzinger-orig-water-tanks} and real safety-critical systems~\cite{iclp24-pcapump-ec-scasp-shortened}.
We show how to address the identified classes of Zeno-like behaviors so that the non-termination issues 
are avoided.
More precisely, by \emph{Zeno-like} behavior, we mean behavior that makes the reasoning \emph{explore} an infinite number of events with \emph{infinitesimal} time intervals
in between, but an infinite number of events do not \textit{actually occur}, making it only similar to Zeno behavior.
It is caused by common EC modeling patterns, e.g., preventing repeated triggering, some of which are specific to goal-directed reasoning, e.g., terminating specific trajectories.
We have identified a unifying feature of all these behaviors, a \textit{Zeno-descending chain of events} (Section~\ref{sec:zeno-descending-chain}), and propose a technique for 
detecting it during reasoning (Section~\ref{sec:automated-detection}).
We introduce each of the general Zeno-like behaviors using examples from the literature or their variations (Section~\ref{sec:repeated-trigger} and~\ref{sec:trajectory-related}) and we propose solutions for each of them.
Then, we discuss the water tanks example which combines all the 
behaviors
as sub-problems and, in addition, exhibits real Zeno behavior (Section~\ref{sec:real-zeno-problems}).
This example illustrates all the prior solutions combined and solving the Zeno-like sub-problems leads to the real Zeno behavior being dealt with as well.

\secVspaceB{}\section{Background \& Related Work}\secVspaceA{}\label{sec:background} 

Zeno's paradoxes of motion were formulated by the ancient philosopher Zeno of Elea:
his Racetrack paradox (or dichotomy of motion) says that a runner will never reach the finish line because they must run half of the distance first, then half of what is remaining, etc.
%
Zeno executions or Zeno runs are a well-known problem in timed/hybrid systems and automata.
A common definition of Zeno behavior is that an infinite number of events or state transitions 
occur in a finite time interval. 
Zeno behaviors do not occur in real physical systems and are a byproduct of abstractions.
According to~\cite{01-zhang-zeno-hybrid-systems-shortened}, many works simply impose non-Zeno assumptions, such as~\cite{96-henzinger-hybrid-automata}.
A way to deal with Zeno behaviors is to remove them by modifying the system using techniques such as regularization~\cite{99-johansson-bouncing-ball-and-zeno-regularization}. 
Further research proposes techniques for detecting Zeno behavior~\cite{07-timed-automata-zeno-detection, 12-static-detection-of-zeno-runs-shortened},
or sufficient conditions for the presence (or lack of) Zeno behavior~\cite{01-zhang-zeno-hybrid-systems-shortened, 13-lamperski-zeno-stability, 13-coarse-abstractions-make-zeno-difficult-to-detect}.
However, as far as we know, a systematic exploration of Zeno behaviors in EC has not been conducted yet. 
For example, in Section~\ref{sec:repeated-trigger}, we show that the bank account example~\cite{mueller_book-fixed} fundamentally
has no solution in continuous time leading to Zeno-like behavior in goal-directed reasoning.

\ssecVspaceB{}\subsection{s(CASP) and Event Calculus}\ssecVspaceA{}\label{sec:scasp-and-ec}

The s(CASP) system~\cite{scasp-iclp2018-shortened} extends the
expressiveness of ASP systems, based on the stable
model semantics~\cite{gelfond88:stable_models-fixed}, by including
predicates, constraints among non-ground variables, uninterpreted
functions, and, most importantly, a~top-down, query-driven execution
strategy.
This makes it possible to return answers with non-ground
variables 
and to compute
partial models by returning only the necessary fragment of a~stable model
to support the answer, 
which can include the full proof tree making them fully explainable.
In s(CASP), thanks to its constructive negation, %
\code|not p(X)| can return bindings for~\code{X} for which the call
\code{p(X)} would have failed.  Thanks to the interface of s(CASP)
with constraint solvers, sound non-monotonic reasoning with
constraints is possible.
Like other ASP implementations and unlike Prolog, s(CASP) handles
non-stratified negation and returns the corresponding (partial) stable
models, e.g., for \mbox{\code{p :- not q.  q :- not p.}}, under
the stable model semantics there are two possible models for this \textit{even loop},
with either \code{p} or \code{q} being true.

The Event Calculus (EC) is a~formalism for reasoning about events and
change~\cite{shanahan97-book-frame_problem, mueller_book-fixed} 
of which there are several versions. 
We use the Basic Event Calculus (BEC)~\cite{shanahan96-event_calculus-bec-shortened, mueller-ec-chapter}. 
%
There are three basic concepts in EC:
%
(i) an \textit{event} is an action that may occur in the world, 
(ii) a~\textit{fluent} is a~time-varying property of
the world, 
%
(iii) a~\textit{timepoint} is an instant of time.
Events may happen at a~timepoint. Fluents have a~truth value at any
timepoint and may have quantities associated with them.
They change discretely via events or continuously 
via \textit{trajectories}.
An EC formulation of an example consists of a~universal theory, a~domain model, and a narrative.
The theory is a~conjunction of EC axioms, the domain model consists of the causal laws of the domain, 
and the narrative provides observations of event occurrences and properties of fluents. 
Consider the following example from \cite{mueller_book-fixed}:

\begin{figure}[tb]\figVspaceb{}
  \begin{subfigure}[b]{.60\textwidth}
    \begin{subfigure}[b]{.99\textwidth}
        \begin{lstlisting}[style=MySCASP]
fluent(light_on).
event(turn_light_on).   event(turn_light_off).
initiates(turn_light_on,  light_on, T).
terminates(turn_light_off, light_on, T).
% automatically created can_* rules
can_initiates(turn_light_on,  light_on, T).
can_terminates(turn_light_off, light_on, T).
        \end{lstlisting}    
        \vspace{-5pt}\caption{Domain model (\myurlExOne{model.pl}).}
        \label{sfig:ex1-light_on_off-domain}
    \end{subfigure}
    \begin{subfigure}[b]{.99\textwidth}
        \begin{lstlisting}[style=MySCASP]
initiallyN(light_on).
happens(turn_light_on,  10).  happens(turn_light_off, 20).
?- holdsAt(light_on,    T).  % expected: T #> 10,T #=< 20
        \end{lstlisting}    
        \vspace{-7pt}\caption{Narrative and a query with the expected answer.}
        \label{sfig:ex1-light_on_off-narrative}
    \end{subfigure}
  \end{subfigure}
  \begin{subfigure}[b]{.38\textwidth}
    \begin{lstlisting}[style=MySCASP]
holdsAt(Fluent, T2) :-
  T1 .>. 0,   T1 .<. T2,
  can_initiates(Event, Fluent, T1),
  happens(Event, T1),
  initiates(Event, Fluent, T1),
  not_stoppedIn(T1, Fluent, T2).

stoppedIn(T1, Fluent, T2) :-
  T1 .<. T,   T .<. T2,
  can_terminates(Event, Fluent),
  happens(Event, T),
  terminates(Event, Fluent, T).
not_stoppedIn(T1, Fluent, T) :- ...
    \end{lstlisting}
    \vspace{-7pt}\caption{EC axioms at \myurl{axioms/bec_scasp-small.pl}.}
    \label{sfig:ex1-light_on_off-axioms}
  \end{subfigure}
  \vspace{-7pt}\caption[]{s(CASP) encoding of Example~\ref{exa:1_light1} (light on/off).}\figVspaceA{} \vspace{-0.5em} 
  \label{fig:ex1-light_on_off}
\end{figure}

\begin{example}[Light on/off~\linkIconExOne{}]
  \label{exa:1_light1}
  A model of a light that can be turned on and off.
  First, we define the domain model (Fig.~\ref{sfig:ex1-light_on_off-domain}) where:
  (i)~the state of the light is represented by a fluent (line~1),
  (ii)~turning it on/off is represented by events (line~2),
  (iii)~event effects are stated at lines 3--4, and s(CASP) generates lines 6--7 
  (discussed in Section~\ref{sec:ec-axioms-in-scasp}).
  Second, we define a narrative (Fig.~\ref{sfig:ex1-light_on_off-narrative}).
  The light is initially off (line~1) and it is turned on/off at times~10/20 (line~2).
  We query timepoints at which the light is on (line~3) and the answer is between 10 and 20.
  The code is available at \myurlExOne{model.pl}.\footnote{All files are available at \repoExamples{} \artifactDoi{}.}
\end{example}

\noindent
The relevant EC axioms for this example are shown in Fig.~\ref{sfig:ex1-light_on_off-axioms} and implemented in \myurl{axioms/bec_scasp-small.pl}.
The predicate \code{holdsAt(F, T2)} is provable for time \code{T2} if an event \textit{happened} at some earlier time \code{T1} whose effect was to \textit{initiate} that fluent at \code{T1} and the fluent was \textit{not stopped} between \code{T1} and \code{T2}.
A fluent is stopped in a time interval if some event occurs within the interval that terminates it.
The full 7 BEC axioms, used in further examples, are implemented in \myurl{axioms/bec_scasp.pl}. 
Implementation of the predicate \code{not_stoppedIn/3} is discussed in Section~\ref{sec:ec-axioms-in-scasp}.

\sssecVspaceB{}\subsubsection{Modeling EC Axioms in s(CASP)}\sssecVspaceA{}\label{sec:ec-axioms-in-scasp}

Two key factors contribute to the s(CASP)'s ability to model EC in continuous time: the preservation of non-ground variables during the
execution and the integration with constraint solvers.
Using the translation rules, introduced in \cite{arias-ec2022-shortened}, one can translate the BEC axioms into s(CASP) programs that follow
the logic programming convention. 
However, 
special considerations must be made to avoid non-termination. 

First, it is crucial to consider the relative order of \code{happens/2} and \code{initiates/3} (same for \code{terminates/3} and \code{releases/3}) in the axiom for \code{holdsAt/2} shown in Fig.~\ref{sfig:ex1-light_on_off-axioms}.
When proving \code{happens/2} first, the reasoning might go through an expensive (or non-terminating) proof of an event happening only to then realize that it has no effects on the fluent. 
However, proving \code{initiates/3} first can lead to the reasoning trying to prove the effect of an infinite number of events without considering whether they happen\footnote{%
Such non-termination is shown in~\myurlExOne{logs/non_term_summary.log} using a modified version of the light example.  
}.
In \cite{iclp24-pcapump-ec-scasp-shortened}, we proposed to define predicates \code{can_initiates/3}, \code{can_terminates/3}, \code{can_releases/3}, and \code{can_trajectory/4} via preprocessing 
that is done by s(CASP) automatically when using the \code{--ec} command-line option.
It defines a fact \code{can_initiates(E,F,T)} for each fact and/or rule \code{initiates(E,F,T) :- some_body} (and similarly for others) without duplicates.
Custom rules can be defined manually.
This has proven effective in avoiding non-termination while pruning the search space.

Second, s(CASP) implements the \code{not} keyword by generating \textit{dual rules}.
The generation is uninformed of the semantics of the predicates and, thus, generates more rules than needed including ones that may explore unrealistic combinations of values of continuous time arguments. 
We use a custom implementation of the predicates \code{not_stoppedIn(T1, F, T2)} and \code{not_startedIn/3} in order to avoid non-termination.
The custom implementation (in~\myurlwline{axioms/bec_scasp.pl}{169}) leverages the semantics of the predicates to be more efficient and prevent non-termination.

\secVspaceB{}\section{Zeno-Like Behavior in s(CASP): Zeno-Descending Chain of Events}\secVspaceA{}\label{sec:zeno-descending-chain}

In our experiments, the main 
source of non-termination 
are \emph{triggered events}.
An event is 
triggered
if its occurrences are implied by rules of the domain model.
In continuous time, such trigger rules can lead to an infinite number of occurrences and, thus, to an infinite number of changes to reason about.

We have encountered many examples that share the same pattern in their reasoning trees while non-terminating.
The reasoning gets stuck in a loop through 
event trigger rules, where
each iteration 
has a \code{happens(Event, TimeN)} predicate with a new time variable which is strictly less than the one from the previous iteration but otherwise both are constrained to the same interval. 
The tree thus contains a chain: \code{happens(e,T1)}, \code{T2 < T1}, \code{happens(e,T2)}, ..., \code{TN < }...\code{ < T2 < T1}, \code{happens(e,TN)}.
We call such a sequence a \textit{Zeno-descending chain of events} (or Zeno chain).
A~crucial aspect is that there is no time delta between the \code{T}'s, and so the distance between the timepoints can be infinitesimal.
Reasoning with a Zeno chain results in \textit{exploring} an infinite number of event occurrences in a finite time interval, which matches the definition of Zeno behavior.
However, as we show throughout this paper, a Zeno chain can be encountered by goal-directed reasoning even in examples in which only a finite number of events 
occur.
Therefore, not all Zeno chains represent Zeno behavior but rather only \textit{Zeno-like} behavior.

In Section~\ref{sec:automated-detection}, we propose a technique for automatic detection of a Zeno chain, which allows us to alert users to its presence and to halt the non-terminating reasoning.

\ssecVspaceB{}\subsection{Example(s) of a Zeno-Descending Chain of Events}\ssecVspaceA{}\label{sec:example-zeno-descending-chain}



\begin{example}[Simplified bank account~\linkIconExTwo{}]
  \label{exa:2_bank}
  A \emph{simplified version} of the bank account problem~\cite{mueller_book-fixed}
  reasons about a single bank account and money withdrawal from it.
  If the balance is below the minimum of 1000, then a service fee of 10 will be charged via a triggered event.
  In this version, the service fee can only be charged once \textit{per narrative} (instead of once per month).
  Fig.~\ref{fig:ex2-bank_account} shows the s(CASP) encoding of this example and an example narrative with a~few queries and their expected answers.
  The code is available at \myurlExTwo{model.pl}.
\end{example}

\noindent
All queries from Fig.~\ref{sfig:ex2-bank_account-narrative} non-terminate due to the trigger rule of \code{serviceFee} (Fig.~\ref{sfig:ex2-bank_account-trigger}). 
A~simplified snapshot of the non-terminating reasoning tree 
is shown in Fig.~\ref{fig:ex2-trace}.
The Zeno chain starts at line~6 and continues at line~11.
The reasoning enters a Zeno chain because to prove that the fee happened at some \code{T2}, it must consider that it could have already happened earlier at \code{T3}, but then it must consider an even earlier \code{T4}, etc.
Details and solutions are discussed in Section~\ref{sec:repeated-trigger}.

\begin{figure}[tb]\figVspaceb{}
  \begin{subfigure}[b]{.48\textwidth}
    \begin{lstlisting}[style=MySCASP]
fluent(balance(M)).   fluent(noServiceFeeYet).
event(withdraw(M)).   event(serviceFee).

initiates(withdraw(X), balance(NewB), T) :-
  NewB .=. OldB - X,
  holdsAt(balance(OldB), T).
terminates(withdraw(_), balance(OldB), T) :-
  holdsAt(balance(OldB), T).

terminates(serviceFee, noServiceFeeYet, T).
initiates(serviceFee, balance(NewB), T) :-
  NewB .=. OldB - 10,   % service fee is 10
  holdsAt(balance(OldB), T).
terminates(serviceFee, balance(OldB), T) :-
  holdsAt(balance(OldB), T).
    \end{lstlisting}    
    \vspace{-6pt}\caption{Domain model (\myurlExTwo{model.pl}).}
  \end{subfigure}
  \begin{subfigure}[b]{.50\textwidth}
    \begin{subfigure}[b]{.99\textwidth}
        \begin{lstlisting}[style=MySCASP, firstnumber=16]
happens(serviceFee, T) :-
  holdsAt(noServiceFeeYet, T),
  B .<. 1000,   % min balance is 1000
  holdsAt(balance(B), T).
        \end{lstlisting}
        \vspace{-3pt}\vspace{-5pt}\caption{Problematic event trigger rule.}\vspace{-3pt}
        \label{sfig:ex2-bank_account-trigger}
    \end{subfigure}
    \begin{subfigure}[b]{.99\textwidth}
        \begin{lstlisting}[style=MySCASP]
initiallyP(balance(10000)).
initiallyP(noServiceFeeYet).
happens(withdraw(8000), 10).
happens(withdraw(1500), 20).

?- holdsAt(balance(X),  5).  % 10000
?- holdsAt(balance(X),  15). % 2000
?- holdsAt(balance(X),  25). % 490
?- happens(serviceFee,  T).  % smallest T > 20
        \end{lstlisting}
        \vspace{-6pt}\caption{Narrative and queries with expected answers.}
        \label{sfig:ex2-bank_account-narrative}
    \end{subfigure}
  \end{subfigure}
  \vspace{-5pt}\caption[]{s(CASP) encoding of Example~\ref{exa:2_bank} (simplified bank account).}\figVspaceA{}
  \label{fig:ex2-bank_account}
\end{figure}

\begin{figure}[tb]\figVspaceb{}
    \begin{lstlisting}[style=MySCASP]
happens(serviceFee, {T1~[T1 > 0, T1 =< 100]})                       % T1
  holdsAt(noServiceFeeYet, {T1~[T1 > 0, T1 =< 100]})
    initiallyP(noServiceFeeYet)
    not_stoppedIn(0,noServiceFeeYet, {T1~[T1 > 0, T1 =< 100]})
      {T2~[T2 > 0, T2 < 100]} .<. {T1~[T1 > 0, T1 =< 100]}           % T2 < T1
      happens(serviceFee, {T2~[T2 > 0, T2 < 100]})                  % T2
        holdsAt(noServiceFeeYet, {T2~[T2 > 0, T2 < 100]})
          initiallyP(noServiceFeeYet)
          not_stoppedIn(0,noServiceFeeYet, {T2~[T2 > 0, T2 < 100]})
              {T3~[T3 > 0, T3 < 100]} .<. {T2~[T2 > 0, T2 < 100]}    % T3 < T2
              happens(serviceFee, {T3~[T3 > 0, T3 < 100]})          % T3
    \end{lstlisting}
  \vspace{-10pt}\caption[]{Zeno-descending chain of events in a simplified trace of Ex.~\ref{exa:2_bank} (full in \myurlExTwo{logs/non_term.log})}
  \figVspaceA{}\vspace{-0.6em}
    \label{fig:ex2-trace}
\end{figure}

In the following sections, we present and describe solutions to a selected class of problems that lead to non-termination through a Zeno chain.
Each class of problems is introduced with an example from the literature or a variation thereof.
Table~\ref{fig:tb-examples_summary_problems} summarizes all the
examples, alongside further unnumbered
examples (available at the repository under the folder named
\myurl{other}), and for each example shows which combination of
problems (Zeno-like or Zeno behavior) is present.

\newcommand{\MX}{\large $\ast$} 
\newcommand{\X}{\large $\ast$}
\newcommand{\GX}{\large $\ast$} 
\begin{table}[tb]
  \setlength{\tabcolsep}{4pt}
  \linespread{0.9}
  \centering
  \smaller
  \vspace{-2pt} 
  \caption{Summary of examples and their problem composition.}
  \vspace{-7pt} 
  \begin{tabular}{lcccccc}
    \toprule
    \multirow{2}{*}{\ }
    &\multirow{2}{*}{\vtop{\hbox{\strut Repeated}\hbox{\strut trigger}}}
    &\multicolumn{3}{c}{Trajectory related problems}
    &\multicolumn{2}{c}{Zeno Behaviors}             \\
    \cmidrule(lr){3-5} \cmidrule(lr){6-7}
    &  & Self-ending & Circular & With inequal. & Trivial & Paradoxes \\
    \midrule
    \linkIconExOne{}        Light on/off          \hfill Ex. \ref{exa:1_light1}  &     &     &     &     &     &     \\
    \linkIconExTwo{}        Simpl. bank account   \hfill Ex. \ref{exa:2_bank}    & \MX &     &     &     &     &     \\
    \linkIconExOtherOne{}   Bank account                                         & \GX &     &     &     &     &     \\
    \linkIconExOtherTwo{}   Light trigger off                                    & \GX &     &     &     &     &     \\
    \linkIconExThree{}      Fading light          \hfill Ex. \ref{exa:3_light2}  &     & \MX &     &     &     &     \\
    \linkIconExOtherThree{} Train gate                                           &     & \GX &     &     &     &     \\
    \linkIconExOtherFour{}  Falling object                                       &     & \GX &     &     &     &     \\
    \linkIconExOtherFive{}  Filling vessel                                       &     & \GX &     &     &     &     \\
    \linkIconExFour{}       Pulsing light         \hfill Ex. \ref{exa:4_light3}  &     & \X  & \MX &     &     &     \\
    \linkIconExOtherSix{}   No-loss water tanks                                  &     & \GX & \GX &     &     &     \\
    \linkIconExOtherSeven{} No-loss bouncing ball                                &     & \GX & \GX &     &     &     \\
    \linkIconExFive{}       Simpl. water tanks    \hfill Ex. \ref{exa:5_tanks1}  & \X  & \X  &     & \MX &     &     \\
    \linkIconExApxOne{}     Blinking light        \hfill Ex. \ref{exa:a1_light4} & \X  &     &     &     & \MX &     \\
    \linkIconExSix{}        Water tanks           \hfill Ex. \ref{exa:7_tanks2}  & \X  & \X  & \X  & \X  &     & \MX \\
    \linkIconExOtherEight{} Bouncing ball                                        &     & \GX & \GX &     &     & \GX \\ 
    \bottomrule
  \label{fig:tb-examples_summary_problems}
  \tabVspaceA{}
  \end{tabular}
\end{table}

\ssecVspaceB{}\subsection{Automated Detection of Zeno-Descending Chains of Events}\ssecVspaceA{}\label{sec:automated-detection}

We propose a simple yet effective technique to detect Zeno chains,
made available as a feature of s(CASP) from version \version{} at \url{https://gitlab.software.imdea.org/ciao-lang/sCASP}.  It
can be enabled using the \code{--zeno_halt} flag.  When a Zeno chain
is detected, s(CASP) stops and issues a warning
to inform the user
(avoiding non-termination).
%
This technique,
works as follows:
when a predicate \code{happens(E,T)} is expanded, s(CASP) will check the current derivation looking for predicates \code{happens(E,T1)} and \code{happens(E,T2)}, such as: (i)
all three have the same event~\code{E}, the variables \code{T1} and \code{T2} are constrained to the same interval (\code{low < Tx < high}), (ii) \code{happens(E,T1)} was expanded earlier than \code{happens(E,T2)}, and (iii) \code{T2} is strictly smaller than \code{T1}, i.e., there is a constraint \code{T2 < T1}.
E.g., the Zeno chain shown in Fig.~\ref{fig:ex2-trace} between lines~6 and~11 would be detected 
(see the full detection trace in~\myurlExTwo{logs/zeno_halt.log}).
%
%

We validated the ability of this technique to
(quickly) 
detect Zeno-like behaviors with each of the examples listed in
Table~\ref{fig:tb-examples_summary_problems} (their detection traces are available in the repository).

\secVspaceB{}\vspace{-0.9em}\section{Preventing Repeated Triggering of Events}\secVspaceA{}\vspace{-0.1em}\label{sec:repeated-trigger}

A common concept in EC is preventing events from triggering repeatedly~\cite{mueller_book-fixed}. 
However, reasoning in continuous time introduces issues. 
Consider Ex.~\ref{exa:2_bank} 
from the previous section.
A service fee should be charged if two conditions are met: (i) the balance is below some minimum value and (ii) the fee has not been charged yet.
When money is withdrawn at time \code{T1}, the new balance will be provable via \code{holdsAt(ballance(X),T2)} 
for any \code{T2} such that \code{T2 > T1} (recall the axiom from Fig.~\ref{sfig:ex1-light_on_off-axioms}).
Should the withdrawal make the balance go below the minimum, Condition (i) would hold for all \code{T2}'s (a~continuous interval).
Without Condition (ii), the fee would be charged infinitely many times. 
With Condition (ii), 
the first occurrence \textit{prevents any subsequent} ones.
However, there is a problem when we consider continuous time---it is impossible to find the first occurrence due to the 
non-inclusive lower bound, in our case \code{T2} such that \code{T2 > T1}.
As a consequence, 
it is inherently impossible to answer the query from Fig~\ref{sfig:ex2-bank_account-narrative}
considering continuous time, causing non-termination in s(CASP) due to its attempt to find the smallest \code{T2} (recall Fig.~\ref{fig:ex2-trace}). 
To solve this problem we propose two solutions: 
\textbf{Solution 1 (Add Epsilon)}: An epsilon in time is introduced and the fee is charged after a delay (see Fig.~\ref{sfig:ex2-solutions-1}).
This is achieved using a new predicate \code{holdsAt(F, T, EPS, E)} (see its encoding in Fig.~\ref{sfig:ex2-solutions-axiom}).
The third argument makes the predicate provable only at time \code{T = T1 + EPS} where \code{T1} is the timepoint of occurrence of the event \code{E}, specified by the fourth argument, that initiates the fluent \code{F}.
The fourth argument is not strictly needed but helps avoid a slowdown from exploring irrelevant events. 
This solution ensures that the trigger rule proves \code{happens(withdraw(X), T1)} first and only then triggers the \code{serviceFee} at some \code{T2 = T1 + EPS}.
Since \code{withdraw(X)} is not a triggered event, there only is a finite number of \code{T1}'s to explore. 
Hence, non-termination through this 
rule is no longer possible and the query terminates 
(see \myurlExTwo{logs/fix-holdsAt4.log}).
Introducing an epsilon here is not like discretizing since \code{T1} and \code{T2} can still take any value in the continuous domain (and there is no impact on reasoning time).

\textbf{Solution 2 (Remodel Behavior)}: We remodel the example with continuous time in mind to charge the fee at the same time as the withdrawal and combine the effects of the two events (see Fig.~\ref{sfig:ex2-solutions-2}).
The fee is charged at the originally lower bound instead of infinitely close to it, and the query terminates 
(see \myurlExTwo{logs/fix-remodel.log}).
In this example, the effect of the \code{serviceFee} is to initiate a new balance by combining both the amount subtracted by the fee and 
the event which caused it.
The \code{withdraw(X)} event has no effect if the fee is charged at the same time.
This solution is closest to the original behavior
, but it is only suitable for examples where the effects of the events can be combined. 

\begin{figure}[tb] \figVspaceb{}
  \begin{subfigure}[b]{.45\textwidth}
    \begin{subfigure}[b]{1\textwidth}
        \vspace{0.5pt}
        \begin{lstlisting}[style=MySCASP]
happens(serviceFee, T2) :-
  EPS .=. 1/1000000,   B .<. 1000,
  holdsAt(balance(B), T2, EPS, withdraw(_)),
  holdsAt(noServiceFeeYet, T2).
        \end{lstlisting}    
        \vspace{-5pt}\vspace{-3pt}\caption{Solution 1: Add Epsilon (\myurlExTwo{fix-holdsAt4.pl}).}
        \label{sfig:ex2-solutions-1}
    \end{subfigure}
    \begin{subfigure}[b]{1\textwidth}
        \begin{lstlisting}[style=MySCASP]
holdsAt(Fluent, T2, Dur, Event) :-
  T2 .=. T1 + Dur, ..., happens(Event, T1),
  initiates(Event, Fluent, T1), ...
        \end{lstlisting}    
        \vspace{-5pt}\caption{New axiom (\myurlwline{axioms/bec_scasp.pl}{118}).}
        \label{sfig:ex2-solutions-axiom}
    \end{subfigure}
  \end{subfigure}
  \begin{subfigure}[b]{.53\textwidth}
    \begin{lstlisting}[style=MySCASP]
happens(serviceFee, T) :- happens(withdraw(Amount), T),
  holdsAt(noServiceFeeYet, T),  NewB .<. 1000,
  NewB .=. OldB - Amount, holdsAt(balance(OldB), T).
initiates(withdraw(X), balance(NewB), T) :-
  not_happens(serviceFee, T), ...
terminates(withdraw(_), balance(OldB), T) :-
  not_happens(serviceFee, T), ...
initiates(serviceFee, balance(NewB), T) :-
  happens(withdraw(X), T),
  NewB .=. (OldB - X) - 10, holdsAt(balance(OldB), T).
    \end{lstlisting}
    \vspace{-5pt}\caption{Solution 2: Remodel Behavior (\myurlExTwo{fix-remodel.pl}).}
    \label{sfig:ex2-solutions-2}
  \end{subfigure}
  \vspace{-5pt}\caption[]{Two solutions to the problem in Example~\ref{exa:2_bank} (simplified bank account).}\figVspaceA{}
  \label{fig:ex2-solutions}
\end{figure}

Using the above solutions, we modeled additional examples that feature the same problem:
(i)~the \textbf{full bank account
\linkIconExOtherOne{}} models multiple accounts and transactions and adds a monthly reset for the service fee (encoding in \myurl{other/1-bank_account}).
(ii)~the \textbf{light trigger
off~\linkIconExOtherTwo{}}, a version of Ex.~\ref{exa:1_light1}
where the light automatically turns off when it is on (encoding in \myurl{other/2-light_trigger_off}).

\secVspaceB{}\section{Trajectory-Related Problems}\secVspaceA{}\label{sec:trajectory-related}

EC represents continuous change via \textit{trajectories}.
A trajectory, 
defined using \code{trajectory(F1,T1,F2, T2) :- body}, starts when its \textit{control fluent} \code{F1} is initiated at \code{T1} by an event and stays active until the fluent is terminated.
While the trajectory is active, it is used to determine the value of the \textit{continuous fluent} \code{F2} at any \code{T2}.
%
Below, we discuss three non-termination problems with trajectories that we have encountered.

\ssecVspaceB{}\subsection{Self-Ending Trajectory Problem}\ssecVspaceA{}\label{sec:self-ending-trajectory}

Autotermination is a common modeling pattern where a trajectory ends via an event triggered based on the continuous value defined by the trajectory. 
We call such trajectories \textit{self-ending}. 
Modeling these in s(CASP) can lead to non-termination.
Consider the following example:

\begin{example}[Fading light~\linkIconExThree{}, cont. Ex. \ref{exa:1_light1}]
  \label{exa:3_light2}
  When the light is turned on, its brightness gradually fades in from 0 until it reaches full brightness of 10. 
  The continuous change of the fluent \code{brightness(X)} is modeled by a trajectory controlled by the fluent \code{fading_in} and ends automatically via a triggered event \code{fade_in_end}.
  The event triggers when full brightness is reached
  and
  initiates \code{brightness(10)} as the new discrete value of the fluent.
  Fig.~\ref{fig:ex3-fading_light} shows the new parts of the s(CASP) encoding (compared to Ex. \ref{exa:1_light1}) and an example narrative with a~few queries.
  The code is available at \myurlExThree{model.pl}.
\end{example}

\noindent
All queries in Fig.~\ref{sfig:ex3-fading_light-narrative} non-terminate, looping through the trigger rule (Fig.~\ref{sfig:ex3-fading_light-model}, line 13) and the trajectory axiom (Fig.~\ref{sfig:ex3-fading_light-axiom}, lines 1-4).
The Zeno chain looks like this: \code{happens(fade_in_end,T1)}, \code{holds- At(brightness(10),T1)}, 
\code{T2<T1}, \code{can_initiates(fade_in_end,brightness(10),T2)}, \code{happens(fa- de_in_end,T2)} 
(see \myurlExThree{logs/non_term_summary.log}). This chain can be detected automatically (see~\myurlExThree{logs/zeno_halt.log}).
The problem is that the effect of the event is considered as a way to prove its own trigger. 
This is a limitation of goal-directed reasoning as the example does not inherently contain any Zeno(-like) behavior.
We propose one~solution:
%
%

\begin{figure}[tb]\figVspaceb{}
  \begin{subfigure}[b]{.54\textwidth}
    \begin{lstlisting}[style=MySCASP]
event(fade_in_end).   fluent(light_on).
fluent(brightness(X)).   fluent(fading_in).
initiates(turn_light_on, fading_in, T).
releases(turn_light_on, brightness(X), T).
terminates(fade_in_end, fading_in, T).
initiates(fade_in_end, brightness(10), T).
terminates(fade_in_end, brightness(X), T):- X .<>. 10.

trajectory(fading_in, T1, brightness(NewB), T2) :-
  NewB .=. OldB + &(&(T2-T1&) * 1&),
  holdsAt(brightness(OldB), T1).
  
happens(fade_in_end, T) :- holdsAt(brightness(10), T).
    \end{lstlisting}    
    \vspace{-5pt}\caption{Domain model additions (\myurlExThree{model.pl}).}
        \label{sfig:ex3-fading_light-model}
  \end{subfigure}
  \begin{subfigure}[b]{.44\textwidth}
    \begin{subfigure}[b]{.99\textwidth}
        \begin{lstlisting}[style=MySCASP]
holdsAt(Fluent2, T2):-  T1 .>. 0,  T1 .<. T2,
  can_trajectory(Fluent1, T1, Fluent2, T2),
  can_initiates(Event, Fluent1, T1),
  happens(Event, T1),
  initiates(Event, Fluent1, T1),
  trajectory(Fluent1, T1, Fluent2, T2),
  not_stoppedIn(T1, Fluent1, T2).
        \end{lstlisting}
        \vspace{-3pt}\vspace{-5pt}\caption{Trajectory axiom (\myurl{axioms/bec_scasp.pl}).}
        \label{sfig:ex3-fading_light-axiom}
    \end{subfigure}
    \begin{subfigure}[b]{.99\textwidth}
        \begin{lstlisting}[style=MySCASP]
initiallyP(brightness(0)).
happens(turn_light_on,    10).
?- holdsAt(brightness(X), 25). % 10
?- happens(fade_in_end,   T).  % 20
        \end{lstlisting}
        \vspace{-5pt}\caption{Narrative and some queries.}
        \label{sfig:ex3-fading_light-narrative}
    \end{subfigure}
  \end{subfigure}
  \vspace{-5pt}\caption[]{s(CASP) encoding of Example~\ref{exa:3_light2} (fading light), extension of Example~\ref{exa:1_light1} (light on/off)}\figVspaceA{}
  \label{fig:ex3-fading_light}
\end{figure}%

\begin{figure}[tb]\figVspaceb{}
  \begin{subfigure}[b]{.50\textwidth}
        \begin{lstlisting}[style=MySCASP]
happens(fade_in_end, T) :-
  holdsAt(brightness(10), T, fading_in).
        \end{lstlisting}
        \vspace{-5pt}\caption{Solution: Restrict HoldsAt (\myurlExThree{fix-holdsAt3.pl}).}
        \label{sfig:ex3-solutions-solution}
  \end{subfigure}
  \begin{subfigure}[b]{.48\textwidth}
        \begin{lstlisting}[style=MySCASP]
holdsAt(Fluent2, T2, Fluent1) :- ...,
  trajectory(Fluent1, T1, Fluent2, T2), ...
        \end{lstlisting}
        \vspace{-5pt}\caption{New axiom (\myurlwline{axioms/bec_scasp.pl}{51}).}
        \label{sfig:ex3-solutions-axiom}
    \end{subfigure}
  \vspace{-5pt}\caption[]{Solution for the problem in Example~\ref{exa:3_light2} (fading light).}\figVspaceA{}\vspace{-0.5em}
  \label{fig:ex3-solutions}
\end{figure}

\textbf{Solution 1 (Restrict HoldsAt)}: 
When reasoning about ending a trajectory, we 
assume that it is 
active; otherwise we would not need to reason about its end.
We restrict that \code{holdsAt(brightness(10), T)} should only be proven via the trajectory controlled by the fluent \code{fading_in} (see Fig.~\ref{sfig:ex3-solutions-solution})
using a new predicate \code{holdsAt(F,T,CF)} (Fig.~\ref{sfig:ex3-solutions-axiom}),
which limits the choice of control fluents when proving the value of a fluent. 
This prevents the original loop, and the query terminates 
(see \myurlExThree{logs/fix-holdsAt3.log}).
%
%

The above problem is present in all the trajectory-related examples in the rest of this paper.
The solution from this section is applied in all of them by default. 
In addition, we have modeled the following examples that feature the same problem: 
(i)~the \textbf{filling vessel~\linkIconExOtherFive{}}~\cite{shanahan97-book-frame_problem} which stops filling and starts spilling once it fills up to its maximum level 
(encoding in \myurl{other/5-filling_vessel/}). 
(ii)~The \textbf{train gate~\linkIconExOtherThree{}} controller~\cite{Alur94atheory-shortened} models a railway crossing with trajectories for lowering/raising the crossing gates (encoding in \myurl{other/3-train_gate/}).
(iii)~The \textbf{falling object~\linkIconExOtherFour{}} example~\cite{mueller_book-fixed} models an object that is dropped from some height and falls to the ground 
(encoding in \myurl{other/4-falling_object/}).

\ssecVspaceB{}\subsection{Circular Trajectory Problem}\ssecVspaceA{}\label{sec:circular-trajectory}


This section introduces a problem of \emph{circular dependency} between trajectories---the end of one starts the other and vice versa.
The following example does not inherently contain any Zeno(-like) behavior; however, it is a problem for goal-directed, top-down reasoning in continuous time.

\begin{example}[Pulsing light~\linkIconExFour{}, cont. Ex. \ref{exa:3_light2}]
  \label{exa:4_light3}
  The light fades in and out repeatedly. 
  As soon as the light reaches maximum brightness, it will start fading out back to zero.
  Once zero is reached, the light will start fading in again, repeating the process.
  Fig.~\ref{fig:ex4-pulsing_light} shows the modification of the s(CASP) encoding (compared to Ex. \ref{exa:3_light2}) and a narrative with a~few queries.
  The code is available at \myurlExFour{model.pl}.
\end{example}

\begin{figure}[t] 
  \begin{subfigure}[b]{.57\textwidth}
    \begin{lstlisting}[style=MySCASP]
fluent(fading_out).         event(fade_out_end).
terminates(fade_out_end, fading_out, T).
initiates(fade_out_end, fading_in, T). 
trajectory(fading_out, T1, brightness(NewB), T2) :-
  NewB .=. OldB - &(T2-T1&), holdsAt(brightness(OldB), T1).
happens(fade_out_end, T) :-
  holdsAt(brightness(0), T, fading_out).
    \end{lstlisting}    
    \vspace{-5pt}\caption{Encoding additions (\myurlExFour{model.pl}).}
    \label{sfig:ex4-pulsing_light-model}
  \end{subfigure}
  \begin{subfigure}[b]{.41\textwidth}
    \begin{lstlisting}[style=MySCASP]
initiallyP(brightness(0)).
happens(turn_light_on,   10).

?- happens(fade_in_end,  T). % 20
                             % and 40
?- happens(fade_out_end, T). % 30
    \end{lstlisting}
    \vspace{-5pt}\caption{Narrative and some queries.}
    \label{sfig:ex4-pulsing_light-narrative}
  \end{subfigure}
  \vspace{-5pt}\caption[]{s(CASP) encoding of Example~\ref{exa:4_light3} (pulsing light), extension of Example~\ref{exa:3_light2} (fading light).}\figVspaceA{}\vspace{-0.1em}
  \label{fig:ex4-pulsing_light}
\end{figure}

\begin{figure}[tb] \figVspaceb{}
  \begin{subfigure}[b]{.49\textwidth}
    \begin{subfigure}[b]{.99\textwidth}
        \begin{lstlisting}[style=MySCASP]
happens(fade_in_end, T) :-   Dur .=. 10,
  holdsAt(brightness(10), T, fading_in, Dur).
happens(fade_out_end, T) :-   Dur .=. 10,
  holdsAt(brightness(0), T, fading_out, Dur).
        \end{lstlisting}    
        \vspace{-3pt}\vspace{-5pt}\caption{Solution 1: Add Duration (\myurlExFour{fix-holdsAt4.pl}).}
        \label{sfig:ex4-solutions-solution1}
    \end{subfigure}
    \begin{subfigure}[b]{.99\textwidth}
        \begin{lstlisting}[style=MySCASP]
holdsAt(Fluent2, T2, Fluent1, Duration) :-
  T2 .=. T1 + Duration, ...
  trajectory(Fluent1, T1, Fluent2, T2), ...
        \end{lstlisting}
        \vspace{-5pt}\caption{New axiom (\myurlwline{axioms/bec_scasp.pl}{65}).}
        \label{sfig:ex4-solutions-axiom}
    \end{subfigure}
  \end{subfigure}
  \begin{subfigure}[b]{.49\textwidth}
    \begin{subfigure}[b]{.99\textwidth}
        \begin{lstlisting}[style=MySCASP]
incr_event(fade_in_end).
incr_event(fade_out_end).

can_initiates(fade_in_end, fading_out, T) :-
  incr_happens(fade_in_end, T).
  
can_initiates(fade_out_end, fading_in, T) :-
  incr_happens(fade_out_end, T).
        \end{lstlisting}
        \vspace{-5pt}\caption{Solution 2: Incremental Reasoning (\myurlExFour{fix-incr.pl}).}
        \label{sfig:ex4-solutions-solution2}
    \end{subfigure}
  \end{subfigure}
  \vspace{-7pt}\caption[]{Two solutions for the problem in Example~\ref{exa:4_light3} (pulsing light).}\figVspaceA{}\vspace{-1em}
  \label{fig:ex4-solutions}
\end{figure}

\noindent
All the queries from Fig.~\ref{sfig:ex4-pulsing_light-narrative} non-terminate.
To prove the effect of a trajectory at some~\code{T2}, we first need to find its start time~\code{T1} (see the axiom in Fig.~\ref{sfig:ex3-fading_light-axiom}).
However, due to the circularity, the reasoning will proceed as follows:
(i)~try to prove the effect of trajectory~\code{A} at~\code{TA2},
(ii)~need to prove its start time~\code{TA1} such that \code{TA1 < TA2},
(iii)~trajectory~\code{A} can start via the end of an earlier trajectory~\code{B},
(iv)~need to prove the effect of trajectory~\code{B} at \code{TB2} such that \code{TB2 = TA1}, which is Step~(i) with \code{TB2 < TA2}. 
The loop contains the 
trajectory axiom (Fig.~\ref{sfig:ex3-fading_light-axiom} lines 1--4) and the trigger rules of the trajectory start/end events (Fig.~\ref{sfig:ex4-pulsing_light-model} lines 6--7, Fig.~\ref{sfig:ex3-solutions-solution} lines 1-2).
The Zeno chain looks as follows: \code{happens(fade_out_end,T1)}, \code{holdsAt(brightness(0),T1)}, \code{T2<T1}, 
\code{happens(fade_in_end,T2)}, \code{holdsAt(brightness(10),T2)}, \code{happens(fade_out_end,T3)} (see \myurlExFour{logs/non_term_summary.log}); and it can be automatically detected (see~\myurlExFour{logs/zeno_halt.log}).
The problem is that the reasoning never reaches the body of the \code{trajectory/4} predicate which means that there is no knowledge of the potential duration of the trajectory, i.e., the reasoning is exploring an 
infinite number of infinitely short trajectories. 
We propose two solutions for this problem:

\textbf{Solution 1 (Add Duration)}: Introduces duration information into the event trigger rules (see Fig.~\ref{sfig:ex4-solutions-solution1}) using 
a new predicate \code{holdsAt(F1,T,F2,Duration)} (see Fig.~\ref{sfig:ex4-solutions-axiom}) 
which is based on the \code{holdsAt/3} predicate from Ex.~\ref{exa:3_light2} 
and extended with a duration similarly to the \code{holdsAt/4} 
predicate from Ex.~\ref{exa:2_bank}. 
It specifies a trajectory control fluent~\code{F2} and a \code{Duration} for which the fluent must hold.
This introduces the duration of the trajectories into the 
loop and the query terminates 
(see \myurlExFour{logs/fix-holdsAt4.log}).
This approach is suitable for cases where the trajectories always have the same duration; for cases with variable duration, we
could specify a minimum duration or use the next solution.

\textbf{Solution 2 (Incremental Reasoning)}: Uses 
forward reasoning for trajectories 
and accurately deals with trajectories with variable duration.
The reasoning 
starts with the \textit{first trajectory}. 
Once we prove its end, we can start reasoning whether it caused the start of some second trajectory and so on.
This is achieved by taking the trajectory ending events (Fig.~\ref{sfig:ex4-solutions-solution2}, lines~1-2) and decoupling all their effects, other than terminating the trajectory, using \code{incr_happens(E, T)} (Fig.~\ref{sfig:ex4-solutions-solution2}, lines~4-8).
The 
reasoning repeatedly executes queries to see if any of the 
events happen.
If an occurrence of 
\code{E1} is found at 
\code{T1}, then a fact \code{incr_happens(E1, T1)}
is added to the knowledge base.
The next increment will then query events at times greater than \code{T1}.
If multiple occurrences are found in an increment, only the earliest ones are saved as they could impact the occurrence of the later ones in the next increment. 
This process is repeated until no more occurrences 
are found or until a 
time limit is reached.
The original query is then executed (see \myurlExFour{logs/fix-incr.log}) with the knowledge base of occurrences of all incremental events. 
This approach is implemented in s(CASP) and can be enabled via the \code{--incremental} cmd-line option.
%


Using the above, we have modeled two additional examples: 
(i)~The \textbf{no-loss bouncing ball~\linkIconExOtherSeven{}} is a simplified bouncing ball that 
bounces up to the original height on every bounce (encoding in
\myurl{other/7-no_loss_bouncing_ball}).
(ii)~The \textbf{no-loss water tanks~\linkIconExOtherSix{}} models two
water tanks which are being slowly drained and one at a time is being
refilled by a water input arm
, while only
considering narratives where the combined draining rate is the same as
the input rate (encoding in \myurl{other/6-no_loss_water_tanks}).

\ssecVspaceB{}\subsection{Self-Ending Trajectory with End Condition Based on Inequality}\ssecVspaceA{}\label{sec:self-ending-trajectory-ineq}

In the previous trajectory examples, we considered 
trajectories whose end events were triggered based on equality, e.g., the light stops fading in when its brightness is \textit{equal} to 10.
In the case of strictly monotonic trajectories, 
the brightness will be exactly 10 only at one timepoint, and so the end event can only occur once. 
However, 
sometimes an end condition with inequality may be called for.
In such cases, we can face a manifestation of the repeated trigger problem from Ex.~\ref{exa:2_bank} 
where 
an event was defined to occur infinitely close to some non-inclusive bound. 
Consider the following example:

\begin{figure}[tb]\figVspaceb{}
  \begin{subfigure}[b]{.52\textwidth}
    \begin{lstlisting}[style=MySCASP]
fluent(water_left(X)).
fluent(left_filling).   fluent(left_draining).
event(switch_left).   event(switch_right).

initiates(switch_left, left_filling, T).
terminates(switch_left, left_draining, T).
terminates(switch_right, left_filling, T).
initiates(switch_right, left_draining, T).

trajectory(left_filling, T1, water_left(NewW), T2) :-
  NewW .=. OldW + ((T2-T1) * 10), % rate 30 - 20
  holdsAt(water_left(OldW), T1).
trajectory(left_draining, T1, water_left(NewW), T2) :-
  NewW .=. OldW - ((T2-T1) * 20), % out rate 20
  holdsAt(water_left(OldW), T1).

happens(switch_left, T):- W .=<. 50, % target lvl 50
  holdsAt(water_left(W), T, left_draining).
    \end{lstlisting}    
    \vspace{-5pt}\caption{Partial domain model (full at \myurlExFive{model.pl}).}
    \label{sfig:ex5-simplified_water_tanks-model}
  \end{subfigure}
  \begin{subfigure}[b]{.46\textwidth}
    \begin{subfigure}[b]{.99\textwidth}
        \begin{lstlisting}[style=MySCASP]
initiallyP(water_left(100)).
happens(start(right),     10).
happens(switch_right,     16.25).

?- happens(switch_left,   T). % 12.5, 18.125
?- holdsAt(water_left(X), 12.5).  % 50
?- holdsAt(water_left(X), 16.25). % 87.5
?- holdsAt(water_left(X), 18.125).% 50
?- holdsAt(water_left(X), 19.5).  % 50.625
        \end{lstlisting}
        \vspace{-3pt}\vspace{-5pt}\caption{Working narrative (\myurlExFive{model-nar1.pl}).}
        \label{sfig:ex5-simplified_water_tanks-narrative1}
    \end{subfigure}
    \begin{subfigure}[b]{.99\textwidth}
        \begin{lstlisting}[style=MySCASP]
initiallyP(water_left(0)).
happens(start(left),      10).
happens(switch_right,     13).

?- holdsAt(water_left(X), 13).% 30 (term.)
?- happens(switch_left, T).   % smallest T>13
?- holdsAt(water_left(X), 15).% 50
        \end{lstlisting}
        \vspace{-5pt}\caption{Non-term. narrative (\myurlExFive{model-nar2.pl}).}
        \label{sfig:ex5-simplified_water_tanks-narrative2}
    \end{subfigure}
  \end{subfigure}

  \vspace{-5pt}\caption[]{s(CASP) encoding of Example~\ref{exa:5_tanks1} (simplified water tanks).}\figVspaceA{}\vspace{-0.3em}
  \label{fig:ex5-simplified_water_tanks}
\end{figure}

\begin{figure}[tb]\figVspaceb{}
  \begin{subfigure}[b]{.52\textwidth}
    \begin{lstlisting}[style=MySCASP]
happens(switch_left, T) :- W .=. 50,  % = target lvl
  holdsAt(water_left(W), T, left_draining).
happens(switch_left, T) :-   Dur .=. 1,
  W .<. 50,   % < target lvl
  holdsAt(water_left(W), T, left_draining, Dur).
    \end{lstlisting}
    \vspace{-5pt}\caption{Solution 1: Add Duration (\myurlExFive{fix-split_holdsAt4.pl}).}
        \label{sfig:ex5-solutions-solution1}
  \end{subfigure}
  \begin{subfigure}[b]{.46\textwidth}
    \begin{lstlisting}[style=MySCASP]
terminates(switch_right, left_filling, T) :-
  X .>. 50, % > target level 50
  holdsAt(water_left(X), T, left_filling).
initiates(switch_right, left_draining, T) :-
  X .>. 50, % > target level 50
  holdsAt(water_left(X), T, left_filling).
    \end{lstlisting}    
    \vspace{-5pt}\caption{Solution 2: Remodel Beh. (\myurlExFive{fix-split_no_start.pl}).}
    \label{sfig:ex5-solutions-solution2}
  \end{subfigure}
  \vspace{-5pt}\caption[]{Two solutions for the problem in Example~\ref{exa:5_tanks1} (simplified water tanks).}\figVspaceA{}\vspace{-0.8em}
  \label{fig:ex5-solutions}
\end{figure}

\begin{example}[Simplified water tanks~\linkIconExFive{}] 
  \label{exa:5_tanks1}
  A simplified version of the well-known example of two leaking water tanks~\cite{97-alur-henzinger-orig-water-tanks},
  where the tanks share one input arm that switches back and forth between them in an attempt to keep their water levels above a target level.
  Our simplified version only models the left tank, and any switches to the right tank come only as facts in the narrative. 
  There is a trajectory for draining the left tank via a~draining rate (20\,units)
  and a second
  trajectory for filling it via a combination of an input rate (30\,units)
  and the draining rate. 
  A triggered event \code{switch_left} causes the input arm to switch to the left tank. 
  It is triggered while the tank is draining if the current water level is \textit{equal to or below} the target level (50\,units)---the draining trajectory has an end condition with inequality.
  Fig.~\ref{fig:ex5-simplified_water_tanks} shows most of the s(CASP) encoding of this example
  and two example narratives with a~few queries.
  The code is available at \myurlExFive{model.pl}.
\end{example}

\noindent
Queries for the first narrative (Fig.~\ref{sfig:ex5-simplified_water_tanks-narrative1}) all terminate.\footnote{%
However, a more complex impl. of \code{not_stoppedIn/3} is needed (in~\myurlwline{axioms/bec_scasp-interval_not.pl}{194}).%
}
However, the second and third query in the second narrative (Fig.~\ref{sfig:ex5-simplified_water_tanks-narrative2}) do not.
In the first narrative, the arm starts in the right tank meaning that
the tank is initially draining from its starting level~100,
drops to the target level~50
at time~12.5, and a \code{switch_left} is triggered.
A \code{switch_right} happens at time~16.25, when the tank is already filled up to~87.5
(above the target level).
In the second narrative, the tank starts empty and is filling. 
A \code{switch_right} happens at time~13 while the left tank is only at level~30---still below the target level.
The filling trajectory ends at time~13, but a \code{switch_left} is triggered as soon as possible at some \code{T > 13}.
However, this is a problem because an infinitely short draining trajectory is created.
The reasoning non-terminates due to its attempt to find an event infinitely close to a non-inclusive lower bound (see \myurlExFive{logs/non_term_summary.log} and detection in \myurlExFive{logs/zeno_halt.log}).
Such an event cannot be found in continuous time, hence, it is impossible to answer the query. 
This is a manifestation of the repeated trigger problem from Ex.~\ref{exa:2_bank} 
in the context of a trajectory.
We propose two solutions, similar to the ones from Ex.~\ref{exa:2_bank}.
%
%
%
Both solutions split the reasoning 
into two situations: (a)~the trajectory starts when its end condition is not met yet (1st narrative) and (b)~the trajectory starts when its end condition already is met (2nd narrative).
In~Case~(a), the trajectory reaches the equality first and ends before the inequality becomes applicable
, therefore, the inequality is not needed.
Similarly, in Case~(b), the equality is never reached (it has been already exceeded) and hence not needed.

\textbf{Solution 1 (Add Duration)}: Uses an equality-based trigger condition for Case~(a) and introduces a 
minimum duration for the trajectory in Case~(b) using the \code{holdsAt/4} predicate from Ex.~\ref{exa:4_light3} (see Fig.~\ref{sfig:ex5-solutions-solution1}). 
The fixed duration prevents the occurrence of an infinitely short trajectory, removing the non-termination problem 
(see \myurlExFive{logs/fix-split_holdsAt4.log}).
The duration is inherently required by this example and does not lead to the inaccuracies that would arise if we discretized.

\textbf{Solution 2 (Remodel Behavior)}: Also uses an equality-based trigger condition for Case~(a) but a different approach for Case~(b).
An infinitely short trajectory only makes an infinitely small difference, thus, should not start at all.
This can be achieved by limiting
all the effects of the \code{switch_right} event only to Case~(a) (see Fig.~\ref{sfig:ex5-solutions-solution2}) 
and the query terminates 
(see \myurlExFive{logs/fix-split_no_start.log}).

%
%
%

\secVspaceB{}\vspace{0.5em}\section{True Zeno Behavior}\secVspaceA{}\label{sec:real-zeno-problems}

%
We say the prior examples contain \textit{Zeno-like} behaviors as they do not
fully meet the definition of Zeno behavior---the reasoning was \textit{considering} an infinite number of events in finite
time, but they never \textit{needed to happen}.
This section presents a true Zeno behavior in the
\textbf{water tanks~\linkIconExSix{}}~\cite{97-alur-henzinger-orig-water-tanks} where an
infinite number of events occurs due to delays between them getting
shorter and shorter.
In addition, we modeled two more examples:
(i)~the \textbf{bouncing ball~\linkIconExOtherEight{}}~\cite{99-johansson-bouncing-ball-and-zeno-regularization} which looses energy with each bounce.
(ii)~The \textbf{blinking light~\linkIconExApxOne{}}
which blinks 
with an infinitesimal delay, showing trivial Zeno behavior 
(detailed in 
\ref{apx:zero-delay-circular}).

\begin{example}[Water tanks~\linkIconExSix{}, cont. Ex.~\ref{exa:5_tanks1}]
  \label{exa:7_tanks2}
  The full version of the water tanks~\cite{97-alur-henzinger-orig-water-tanks}, where
  the tanks and the arm switches are modeled via triggered events based on reasoning about both water levels.
  The Zeno behavior~\cite{97-alur-henzinger-orig-water-tanks, 01-zhang-zeno-hybrid-systems-shortened, 99-johansson-bouncing-ball-and-zeno-regularization} manifests 
  when the input flow rate is higher than the output flow rates of each tank but lower than their combined output rates---the arm is able to fill each tank, 
  but overall the amount of water in the system is decreasing.
  The arm switches instantaneously and, thus, will keep both tanks from going below the target level by switching for shorter and shorter durations as they both drop closer to their target levels.
  Fig.~\ref{fig:ex7-water_tanks} sketches s(CASP) encoding
  (full code in \myurlExSix{model.pl}).
\end{example}

\noindent
%
 Queries shown in Fig.~\ref{sfig:ex7-water_tanks-narrative} non-terminate  (see \myurlExSix{logs/non_term_summary.log})
due to a combination of the problems that we have presented so far: (a)~real Zeno behavior, (b) the repeated trigger problem from Ex.~\ref{exa:2_bank} manifested in a trajectory, (c) the self-ending trajectory problem from Ex.~\ref{exa:3_light2}, (d) the circular trajectory problem from Ex.~\ref{exa:4_light3}, and (e) the inequality end condition problem from Ex.~\ref{exa:5_tanks1}.
All the sub-problems, except (a), cause non-termination through a Zeno chain and, thus, can be detected (\myurlExSix{logs/zeno_halt.log}).
%
%
%
%
%
Like the two prior examples, the initial encoding already solves 
Problem~(c) via the \code{holdsAt/3} predicate.
We further propose three solutions:

\begin{figure}[tb] \figVspaceb{}
  \begin{subfigure}[b]{.54\textwidth}
    \begin{lstlisting}[style=MySCASP]
fluent(right_filling). % renamed fluent(left_draining)
fluent(water_right(X)).

trajectory(right_filling, T1, water_right(NewW), T2) :-
  ... % similar to trajectory for water_left
trajectory(right_filling, T1, water_right(NewW), T2) :-
  ... % similar to trajectory for water_left
  
happens(switch_right, T) :-
  CurrW .=<. 50,   % target level
  holdsAt(water_right(CurrW), T, left_filling).
    \end{lstlisting}    
    \vspace{-5pt}\caption{Domain model additions (\myurlExSix{model.pl}).}
    \label{sfig:ex7-water_tanks-model}
  \end{subfigure}
  \begin{subfigure}[b]{.44\textwidth}
    \begin{lstlisting}[style=MySCASP]
initiallyP(water_left(100)).
initiallyP(water_right(100)).
happens(start(right),      10).
?- T .=<. 19.5, 
   happens(switch_left, T). % 12.5, 18.125
?- T .=<. 19.5,
   happens(switch_right, T).% 16.25, 19.0625
?- holdsAt(water_left(X),  12.5).  % 50
?- holdsAt(water_right(X), 16.25). % 50
?- holdsAt(water_right(X), 19.5).  % 54.375
?- holdsAt(water_left(X),  19.5).  % 50.625
    \end{lstlisting}
    \vspace{-5pt}\caption{Narrative and some queries.}
    \label{sfig:ex7-water_tanks-narrative}
  \end{subfigure}
  \vspace{-5pt}\caption[]{s(CASP) encoding of Example~\ref{exa:7_tanks2} (water tanks), full version of Example~\ref{exa:5_tanks1}.}
  \figVspaceA{}\vspace{-0.3em}
  \label{fig:ex7-water_tanks}
\end{figure}

\begin{figure}[tb] \figVspaceb{} 
  \begin{subfigure}[b]{.52\textwidth}
    \begin{subfigure}[b]{.99\textwidth}
        \begin{lstlisting}[style=MySCASP]
incr_event(switch_left).   incr_event(switch_right).
can_initiates(switch_left, left_filling, T) :-
  incr_happens(switch_left, T).
can_initiates(switch_right, right_filling, T) :-
  incr_happens(switch_right, T).
?- !incr_max_time(19.5), ...
        \end{lstlisting}    
        \vspace{-3pt}\vspace{-5pt}\caption{Partial Solution 1: Incremental (\myurlExSix{partfix-incr.pl}).}
        \label{sfig:ex7-solutions-solution1}
    \end{subfigure}
    \begin{subfigure}[b]{.99\textwidth}
        \begin{lstlisting}[style=MySCASP]
happens(switch_right, T) :-    MinD .=. 1,
  D .>=. MinD,    W .=<. 50,   % target level 50
  holdsAt(water_right(W), T, left_filling, D).
% same for happens(switch_left, T) :- ...
        \end{lstlisting}
        \vspace{-5pt}\caption{Solution 2: Add Duration(\myurlExSix{fix-holdsAt4.pl}).}
        \label{sfig:ex7-solutions-solution2}
    \end{subfigure}
  \end{subfigure}
  \begin{subfigure}[b]{.46\textwidth}
    \begin{lstlisting}[style=MySCASP]
% same as Solution 1
incr_event(...).  
can_initiates(...) :- incr_happens(...).

% similar to Sol 2., but split trigger rules
happens(switch_right, T) :-   MinD .=. 1,
  D .>=. MinD,   W .=. 50,   % target level 50
  holdsAt(water_right(W), T, left_filling, D).
happens(switch_right, T) :-   D .=. 1, 
  W .<. 50,   % target level 50
  holdsAt(water_right(W), T, left_filling, D).
% same for happens(switch_left, T) :- ...
    \end{lstlisting}
    \vspace{-5pt}\caption{Solution 3: Combined (\myurlExSix{fix-incr_holdsAt4.pl}).}
    \label{sfig:ex7-solutions-solution3}
  \end{subfigure}
  \vspace{-5pt}\caption[]{Three solutions for the problem in Example~\ref{exa:7_tanks2} (water tanks).}\figVspaceA{}\vspace{-0.5em}
  \label{fig:ex7-solutions}
\end{figure}

\textbf{Partial Solution 1 (Incremental Reasoning)}: This solution addresses Problem~(d), which is due to reasoning limitations, and terminates up to Problem~(a).
We solve Problem~(d) via incremental reasoning (see Fig.~\ref{sfig:ex7-solutions-solution1}). 
The incremental events 
are the events for switching the arm left and right. 
Due to the Zeno behavior, typical narratives will not run into Problems~(e) and~(b) as the tanks never go below the target level (cf. \myurlExSix{logs/partfix-incr.log}). 

\textbf{Solution 2 (Add Duration)}: This solution addresses the Zeno-like behavior of Problems~(b) and (e), and the Zeno behavior of Problem~(a).
Problem~(b) manifests as part of Problem~(e), and so we only have to focus on the latter.
We solve Problem~(e) by defining a 
\textit{minimum} duration for the trajectories using the \code{holdsAt/4} predicate (see Fig.~\ref{sfig:ex7-solutions-solution2}).
This introduces a 
delay between the switch events which, additionally, solves Problem~(a)---highlighting the similarity of Zeno and Zeno-like behaviors.    
Moreover, solving Problem~(e) via a minimum duration 
is also an alternative solution to Problem~(d), meaning that incremental reasoning is no longer necessary for termination (see \myurlExSix{logs/fix-holdsAt4.log}).
However, a smaller value of the minimum negatively impacts reasoning time 
as it allows more iterations~of~the~loop.

\textbf{Solution~3 (Combined)}: This solution combines Solutions 1 and 2 (see Fig.~\ref{sfig:ex7-solutions-solution3} and \myurlExSix{logs/fix-incr_holdsAt4.log}). While Solution 1 is only a partial solution, the combination of both solutions leads to better performance (from 30 min to 10 seconds). 

Note that other Zeno behaviors may require still other solutions.
For example, a minimum duration for the trajectory of a \textbf{bouncing ball}~\linkIconExOtherEight{} results in it falling through the floor. 
In this case, a suitable solution is to make it stop bouncing when the last bounce was shorter than a minimum duration or to introduce a flat loss of velocity on each bounce
(modeled in~\myurl{other/8-bouncing_ball}).
%

\secVspaceB{}\section{Conclusion and Summary}\secVspaceA{}\label{sec:conclusion-and-future-work}

\newcommand{\fixx}{{\color{blue} \normalsize \checkmark}}
\newcommand{\term}{{\color{orange} \normalsize \bf~!}}
\begin{table}[tb]
  \addtolength{\tabcolsep}{-0.9pt}
  \linespread{0.9}
  \centering
  \smaller
  \caption{Summary of the proposed solutions and their applicability to the presented problems.}
  \vspace{-7pt} 
  \begin{tabular}{lcccccc}
    \toprule
    \multirow{2}{*}{\ }
    &\multirow{2}{*}{\vtop{\hbox{\strut Repeated}\hbox{\strut trigger}}}
    &\multicolumn{3}{c}{Trajectory related problems}
    &\multicolumn{2}{c}{Zeno Behaviors}
    \\
    \cmidrule(lr){3-5} \cmidrule(lr){6-7}
    &  & Self-ending & Circular & With inequal. & Trivial & Paradoxes  \\
     \midrule
    Automated Detection \hfill (Sect.~\ref{sec:automated-detection}) & \fixx  & \fixx  & \fixx & \fixx & \fixx &     \\[.5em]
    Restrict HoldsAt \hfill (Ex. \ref{exa:3_light2}, \ref{exa:4_light3}, \ref{exa:5_tanks1}, \ref{exa:7_tanks2})            &        & \fixx  &       &       &       &         \\
    Incremental Reasoning \hfill (Ex. \ref{exa:4_light3}, \ref{exa:7_tanks2})      &        &        & \fixx &       &       &          \\
    Add Epsilon / Duration \hfill (Ex. \ref{exa:2_bank}, \ref{exa:4_light3}, \ref{exa:5_tanks1}, \ref{exa:7_tanks2}, \ref{exa:a1_light4})  & \fixx  &        & \fixx & \fixx & \fixx & \fixx    \\
    Remodel Behavior \hfill (Ex. \ref{exa:2_bank}, \ref{exa:5_tanks1})            & \fixx  &        &       & \fixx &       & \fixx   \\
    \bottomrule
  \label{fig:tb-examples_summary_solutions}
  \end{tabular}
  \tabVspaceA{}
\end{table}

We have implemented a novel technique, 
made available as a feature of
s(CASP), that automatically detects Zeno-like behaviors based on the presence of a Zeno-descending chain of events.
We have further identified and characterized a set of non-termination problems, caused by Zeno-like behaviors, that can be encountered in common Event Calculus modeling tasks in continuous time and have proposed four types of solutions that can be used to solve them without discretizing.
    %
    %
    %
We believe that this set of problems (and 
solutions) 
is enough to cover a wide range of real use cases, such as a safety-critical system~\cite{iclp24-pcapump-ec-scasp-shortened}.

As shown in Table~\ref{fig:tb-examples_summary_solutions}, the detection technique works for all classes of problems, except for nontrivial Zeno behaviors because those feature a descending chain of events with delays between them getting shorter and shorter (instead of infinitesimally small).
Additionally, it summarizes the proposed solutions, in which examples they have been applied, and which classes of problems are solved by each of them.
The \textbf{Restrict HoldsAt} solution uses external knowledge of the example to guide EC reasoning by restricting the search rule to chose a specific clause of the predicate \code{holdsAt}. 
\textbf{Incremental Reasoning} addresses circularity via forward reasoning which is particularly suitable for EC due to its need to reconstruct history.
The \textbf{Add Epsilon / Duration} solution introduces a time epsilon into the loop, replacing infinitesimal time steps without causing inaccurate behaviors (as would be the case if we discretized). 
The \textbf{Remodel Behavior} solution tunes the model of the example for reasoning in continuous time to avoid infinitesimal time steps while preserving the intended behavior.
\label{mylastpage}

\bibliographystyle{eptcs}
\bibliography{bibliography,general,clip}%


\newpage
\appendix

\section{Trivial Zeno Behavior: Zero Delay Circular Events}\label{apx:zero-delay-circular}

This appendix presents an example that did not quite fit into the paper.
The example shows trivial Zeno behavior but it is a modeling pattern that is easily encountered in EC.
When reasoning in continuous time, it is easy to introduce event trigger rules which will cause an infinite number of events to occur in a finite time interval.
Events have no duration and their effects are instantaneous, unless we explicitly introduce some duration or delay.
Consider the following example:

\begin{apxExample}[Blinking light, cont. Ex.~\ref{exa:1_light1}]
  \label{exa:a1_light4}
  Both the events for turning the light on and off were changed into triggered events. 
  The trigger rules define that the light should be turned off when it is on and vice versa.
  However, the encoding does not define a frequency for the blinking.
  The triggered events are defined to turn the light off/on \textit{as soon as} it is on/off.
  Fig.~\ref{fig:a1-blinking_light} shows the encoding additions for this example (compared to Ex.~\ref{exa:1_light1}) and an example narrative with a~few queries.
  The code is available at \myurlExApxOne{model.pl}.
\end{apxExample}

\noindent
All the queries in Fig.~\ref{sfig:a1-blinking_light-narrative} will non-terminate and, furthermore, their expected answers are unknown.
There are two problems: (i) two instances of the repeated trigger problem (from Ex.~\ref{exa:2_bank}) one for each triggered event and (ii) trivial zeno behavior since the light is defined to blink infinitely quickly.
Problem (ii) means that while the light is blinking we are unable to conclude whether it is on or off.
The problem manifests as a circular dependency between the two triggered events without any delay in between (see \myurlExApxOne{logs/non_term_summary.log}) and it can be automatically detected (see~\myurlExApxOne{logs/zeno_halt.log}).
\\[-0.7em]

We present our solution in Fig.~\ref{fig:a1-solutions}.
To address Problem (i) we utilize a simplified version of the first solution from Ex.~\ref{exa:2_bank}.
The second one is not suitable here because the turn on/off events have opposite effects.
Solving Problem (i) leads to Problem (ii) being solved as well.


The solution (Fig.~\ref{sfig:a1-solutions-solution}) adds a delay to the event trigger rules using a new predicate \code{holdsAt/3} (Fig~\ref{sfig:a1-solutions-axiom}), which is the same as the \code{holdsAt/4} introduced in Ex.~\ref{exa:2_bank} except without the fourth argument, which is not needed here (there is only one event to consider for each rule).

To address Problem (ii), the Zeno behavior, we need to define some blinking frequency for the light.
Notice that the solution to Problem (i) already introduced a delay between the events and, thus, it solves this problem as well.
This shows the similarity between Zeno behaviors and the Zeno-like behaviors that we have presented in this paper.

\begin{figure}[h]
  \begin{subfigure}[b]{.46\textwidth}
    \begin{lstlisting}[style=MySCASP]
happens(turn_light_off, T) :-
  holdsAt(light_on, T).
happens(turn_light_on, T) :-
  not_holdsAt(light_on, T).
    \end{lstlisting}    
    \vspace{-10pt}\caption{Domain model additions (\myurlExApxOne{model.pl}).}
    \label{sfig:a1-blinking_light-model}
  \end{subfigure}
  \begin{subfigure}[b]{.52\textwidth}
    \begin{lstlisting}[style=MySCASP, numbers=none]
happens(turn_light_on,     10).
?-     holdsAt(light_on,   15). % ? (unknown)
?- happens(turn_light_off, T).  % ? (unknown)
?- happens(turn_light_on,  T).  % 10 and ? (unknown)
    \end{lstlisting}
    \vspace{-10pt}\caption{Narrative and queries with unknown answers.}
    \label{sfig:a1-blinking_light-narrative}
  \end{subfigure}

  \vspace{-7pt}\caption[]{s(CASP) encoding of Example~\ref{exa:a1_light4} (blinking light), based on Example~\ref{exa:1_light1} (light on/off).}
  \label{fig:a1-blinking_light}
\end{figure}

\begin{figure}[!h]
\vspace{-0.8em}
  \begin{subfigure}[b]{.54\textwidth}
    \begin{lstlisting}[style=MySCASP]
happens(turn_light_off, T) :- holdsAt(light_on, T, 1).
happens(turn_light_on, T) :- not_holdsAt(light_on, T, 1).
    \end{lstlisting}    
    \vspace{-10pt}\caption{Solution (\myurlExApxOne{fix-holdsAt3.pl}).}
    \label{sfig:a1-solutions-solution}
  \end{subfigure}
  \begin{subfigure}[b]{.44\textwidth}
    \begin{lstlisting}[style=MySCASP]
holdsAt(Fluent, T2, Dur) :- T2 .=. T1 + Dur,
  ..., happens(Event, T1), ...
    \end{lstlisting}    
    \vspace{-10pt}\caption{New axiom (\myurlwline{axioms/bec_scasp.pl}{80}).}
    \label{sfig:a1-solutions-axiom}
  \end{subfigure}

  \vspace{-7pt}\caption[]{Solution for the problem in Example~\ref{exa:a1_light4} (blinking light).}
  \label{fig:a1-solutions}
\end{figure}

\end{document}